\documentclass[12pt]{article}
\usepackage[dvips]{graphics}

\def\beq   {\begin{equation}}
\def\eeq   {\end{equation}}
\def\beqd  {\begin{displaymath}}
\def\eeqd  {\end{displaymath}}
\def\beqaa {\begin{eqnarray}}
\def\eeqaa {\end{eqnarray}}

\def\noi {\noindent}

\def\ti  {\tilde}

\def\sq  {\ti q}
\def\st  {\ti t}
\def\sb  {\ti b}
\def\sg  {\ti g}

\def\nt  {\tilde\chi^0}
\def\ch  {\tilde\chi^\pm}
\def\chp {\tilde\chi^+}
\def\chm {\tilde\chi^-}

\def\a   {\alpha}
\def\b   {\beta}
\def\t   {\theta}
\def\tst {\theta_{\st}}

\def\sz{\ifmmode{\tilde{\chi}^0} \else{$\tilde{\chi}^0$} \fi}
\def\sw{\ifmmode{\tilde{\chi}} \else{$\tilde{\chi}$} \fi}

\newcommand{\gsim}{\;\raisebox{-0.9ex}
           {$\textstyle\stackrel{\textstyle >}{\sim}$}\;}
\newcommand{\lsim}{\;\raisebox{-0.9ex}{$\textstyle\stackrel{\textstyle<}
           {\sim}$}\;}

\begin{document}
\pagestyle{empty}

\vspace*{-1cm} 
\begin{flushright}
  TGU-31 \\
  UWThPh-2003-14 \\
  ZU-TH 13/03 \\
  hep-ph/0307317
\end{flushright}

\vspace*{1.4cm}

\begin{center}

{\Large {\bf
Impact of CP phases on stop and sbottom searches
}}\\

\vspace{10mm}

{\large 
A.~Bartl$^a$, S.~Hesselbach$^a$, K.~Hidaka$^b$, T.~Kernreiter$^a$ and 
W.~Porod$^c$}

\vspace{6mm}

\begin{tabular}{l}
$^a${\it Institut f\"ur Theoretische Physik, Universit\"at Wien, A-1090
Vienna, Austria}\\
$^b${\it Department of Physics, Tokyo Gakugei University, Koganei,
Tokyo 184--8501, Japan}\\
$^c${\it Institut f\"ur Theoretische Physik, Universit\"at Z\"urich, 
CH-8057 Z\"urich, Switzerland}
\end{tabular}

\end{center}

\vfill

\begin{abstract} 
We study the decays of top squarks ($\tilde t_{1,2}$)
and bottom squarks ($\tilde b_{1,2}$) in the Minimal
Supersymmetric Standard Model (MSSM) with complex
parameters $A_t,A_b,\mu$ and $M_1$. We show that including
the corresponding phases substantially affects
the branching ratios of $\tilde t_{1,2}$ and $\tilde b_{1,2}$
decays in a large domain of the MSSM parameter space.
We find that the branching ratios can easily 
change by a factor of 2 and more when varying the phases.
This could have an important impact on the search for $\st_{1,2}$ 
and $\sb_{1,2}$ and the determination of the MSSM parameters at future 
colliders.
\end{abstract}

\newpage
\pagestyle{plain}
\setcounter{page}{2}


\section{Introduction}

The experimental studies of supersymmetric (SUSY) particles will play
an important role at future colliders. Studying the properties of the
3rd generation sfermions will be particularly interesting because of
the effects of the large Yukawa couplings. The lighter of their mass
eigenstates may be the lightest charged SUSY particles and they
could be investigated at an $e^+ e^-$ linear collider \cite{acco}.
Moreover, they could also be copiously produced in the decays of
heavier SUSY particles.
Several phenomenological studies on SUSY particle searches have been
performed in the Minimal Supersymmetric Standard Model (MSSM)
\cite{ref1} with real SUSY parameters. Analyses of the decays of the
3rd generation sfermions $\tilde{t}_{1,2}$, $\tilde{b}_{1,2}$,
$\tilde{\tau}_{1,2}$ and $\tilde{\nu}_\tau$ in the MSSM with real
parameters have been made in Refs.~\cite{stop2-stau2, stop1}, and
phenomenological studies of production and decays of the 3rd
generation sfermions at future $e^+ e^-$ linear colliders in
Ref.~\cite{rsf}.

In general, however, some of the SUSY parameters may be complex, in
particular the higgsino mass parameter $\mu$, the gaugino mass
parameters $M_{1,2,3}$ and the trilinear scalar coupling parameters
$A_f$ of the sfermions $\tilde{f}$. The SU(2) gaugino mass parameter
$M_2$ can be chosen real after an appropriate redefinition of the
fields.
The experimental upper bounds on the electric dipole moments (EDMs) of
electron, neutron and the $^{199}$Hg and $^{205}$Tl atoms may pose
severe restrictions on the phases of the SUSY parameters, though they 
are model dependent.
In the constrained MSSM the phase of $\mu$ turns out
to be restricted to the range $|\varphi_\mu| \lsim 0.1$ 
if all SUSY masses are in the TeV range \cite{Nath}. 
On the other hand, in more general models, 
such as those with lepton-flavour violation, no restriction on the
phase $\varphi_\mu$ from the electron EDM at one-loop level is obtained
\cite{Bartl:2003ju}. Restrictions arising due to two-loop
contributions to the EDMs are less severe \cite{Pilaftsis}.

Therefore, in a complete phenomenological analysis of production and
decays of the SUSY particles one has to take into account that $A_f$, 
$\mu$ and $M_1$ can be complex. The most direct and unambiguous way to
determine the imaginary parts of the complex SUSY parameters could be
done by measuring relevant CP-violating observables. 
For example, for
the case of sfermion decays CP-violating 
\cite{Aoki:1998cq,Bartl:2002hi} and T-violating \cite{Bartl:2003ck} 
observables have been proposed.
On the other hand, also the CP-conserving observables depend on the
phases of the complex parameters, because in general the
mass-eigenvalues and the couplings of the SUSY particles involved are
functions of the underlying complex parameters. For example, the decay
branching ratios of the Higgs bosons depend strongly on the complex
phases of the $\tilde{t}$ and $\tilde{b}$ sectors 
\cite{Demir:1999hj,ref5',ref4},
while those of the staus $\tilde{\tau}_{1,2}$ and $\tau$-sneutrino
$\tilde{\nu}_\tau$ can be quite sensitive to the phases of the stau
and gaugino-higgsino sectors \cite{CPslepton}.
Also the Yukawa couplings of the third generation sfermions are
sensitive to the SUSY phases at one-loop level \cite{Ibrahim:2003ca}.
Furthermore, explicit CP violation in the Higgs sector can be induced
by $\tilde{t}$ and $\tilde{b}$ loops
if the parameters $A_t$, $A_b$ and $\mu$ are complex
\cite{Demir:1999hj,ref3,ref2}. It is found 
\cite{Demir:1999hj,ref4,ref3,Carena:2002bb} that these loop
effects could significantly influence the phenomenology of the Higgs
boson sector. 

In this article we study the effects of the phases of $A_t$, $A_b$, $\mu$
and $M_1$ on the decay branching ratios of the $\tilde{t}_{1,2}$
and $\tilde{b}_{1,2}$ with $\tilde{q}_1$ ($\tilde{q}_2$) being the
lighter (heavier) squark. We take into account the explicit CP violation
in the Higgs sector. We will show that the influence of the
phases can be quite strong in a large domain of the MSSM parameter
space. This could have an important impact on the search
for $\tilde{t}_{1,2}$ and $\tilde{b}_{1,2}$ and on the determination
of the MSSM parameters at future colliders.

In section 2 we discuss the SUSY CP phase dependences of masses, mixings 
and couplings. Section 3 contains our numerical investigation
on the CP phase dependence of the branching ratios of
$\st_{1,2}$ and $\sb_{1,2}$ decays.
In section 4 we present our conclusions.


\section{SUSY CP Phase Dependences of Masses, Mixings and Couplings} 

In the MSSM the squark sector is specified by the mass matrix in the basis 
$(\sq_L, \sq_R)$ with $\sq=\st$ or $\sb$~\cite{ref7}
\begin{equation}
  {\cal M}^2_{\sq}= 
     \left( 
            \begin{array}{cc} 
                m_{\sq_L}^2 & a_q^* m_q \\
                a_q m_q     & m_{\sq_R}^2
            \end{array} 
     \right)       
                                                           \label{eq:a}
\end{equation}
with
\begin{eqnarray}
  m_{\sq_L}^2 &=& M_{\ti Q}^2 
                  + m_Z^2\cos 2\beta\,(I_3^{q_L} - e_q\sin^2\t_W) 
                  + m_q^2,                                 \label{eq:b} \\
  m_{\sq_R}^2 &=& M_{\{\ti U,\ti D\}}^2  
                  + m_Z^2 \cos 2\b\, e_q\, \sin^2\t_W + m_q^2, 
                                                           \label{eq:c} \\[2mm]
  a_q m_q     &=& \left\{ \begin{array}{l}
                     (A_t - \mu^*\cot\beta) \; m_t~~(\sq=\st)\\
                     (A_b - \mu^*\tan\beta) \; m_b~~(\sq=\sb)
                          \end{array} \right.              \label{eq:d} \\
              &=& \,\, |a_q m_q| \, e^{i\varphi_{\sq}} \,\,
                      (-\pi < \varphi_{\sq} \leq \pi).
                                                           \label{eq:e}
%
\end{eqnarray}
Here $I_3^q$ is the third component of the weak isospin and $e_q$ the 
electric charge of the quark $q$.
$M_{\ti Q,\ti U,\ti D}$ and $A_{t,b}$ are soft SUSY--breaking 
parameters, $\mu$ is the higgsino mass parameter, 
and $\tan\b = v_2/v_1$ with $v_1$ $(v_2)$ being the vacuum 
expectation value of the Higgs field $H_1^0$ $(H_2^0)$. 
As the relative phase $\xi$ between $v_1$ and $v_2$ is irrelevant in our 
analysis, we adopt the $\xi = 0$ scheme \cite{ref3}. 
We take $A_q$ ($q=t,b$) and $\mu$ as complex parameters: 
$A_q = |A_q| \, e^{i\varphi_{A_q}}$ 
and $\mu = |\mu| \, e^{i\varphi_{\mu}}$ with 
$-\pi < \varphi_{A_q,\mu} \leq \pi $. 
Diagonalizing the matrix (\ref{eq:a}) one gets the mass eigenstates $\sq_1$ 
and $\sq_2$
\begin{equation}
     \left( \begin{array}{c} 
                \sq_1 \\
                \sq_2
            \end{array} \right) = R^{\sq} 
     \left( \begin{array}{c} 
                \sq_L \\
                \sq_R
            \end{array} \right) = 
     \left( \begin{array}{cc} 
                e^{i\varphi_{\sq}} \cos\t_{\sq} & \sin\t_{\sq} \\
                -\sin\t_{\sq}  & e^{-i\varphi_{\sq}} \cos\t_{\sq}
            \end{array} \right) 
     \left( \begin{array}{c} 
                \sq_L \\
                \sq_R
            \end{array} \right)
                                                         \label{eq:f}
\end{equation}
with the masses $m_{\sq_1}$ and $m_{\sq_2}$ ($m_{\sq_1} < m_{\sq_2}$), 
and the mixing angle $\t_{\sq}$
\begin{eqnarray}
  m_{\sq_{1,2}}^2 &=& \frac{1}{2}(m_{\sq_L}^2 + m_{\sq_R}^2 
    \mp \sqrt{ (m_{\sq_L}^2 - m_{\sq_R}^2)^2 + 4|a_q m_q|^2 }), 
                                                           \label{eq:g} \\
  \t_{\sq} &=& \tan^{-1}(|a_q m_q|/(m_{\sq_1}^2 - m_{\sq_R}^2)) \quad 
                                         (-\pi/2 \leq \t_{\sq} \leq 0).
                                                           \label{eq:h} 
\end{eqnarray}
The $\sq_L - \sq_R$ mixing is large if $|m_{\sq_L}^2 - m_{\sq_R}^2| 
\lsim |a_q m_q|$, which may be the case in the $\st$ sector due to the large 
$m_t$ and in the $\sb$ sector for large $\tan\b$ and $|\mu|$.
 From Eqs. (\ref{eq:d}),(\ref{eq:g}) and (\ref{eq:h}) we see that 
$m_{\sq_{1,2}}^2$ and $\t_{\sq}$ depend on the phases only through a term 
$\cos(\varphi_{A_q} + \varphi_{\mu})$. This phase dependence of the 
$\st$ ($\sb$) 
sector is strongest if $|A_t| \simeq |\mu| \cot\b$ ($|A_b| 
\simeq |\mu| \tan\b$). 

The properties of charginos $\ch_i$ ($i=1,2$; $m_{\ch_1}<m_{\ch_2}$) and 
neutralinos $\nt_j$ ($j=1,...,4$; $m_{\nt_1}< ... < m_{\nt_4}$) are 
determined by the parameters $M_2, M_1, \mu$ and $\tan\b$.
We assume that the gluino mass $m_{\sg}$ is real. 
We write the U(1) gaugino mass $M_1$ as 
$M_1 = |M_1| e^{i\varphi_1} \, (-\pi < \varphi_1 \leq \pi)$. 
Inspired by the gaugino mass unification we take 
$|M_1| = (5/3) \tan^2\t_W M_2$ and $m_{\sg} = (\a_s(m_{\sg})/\a_2) M_2$. 
In the MSSM Higgs sector with explicit CP violation the 
mass-eigenvalues and couplings of the neutral and charged Higgs bosons, 
$H_1^0, H_2^0, H_3^0$ $(m_{H_1^0} < m_{H_2^0} < m_{H_3^0})$ and $H^\pm$, 
including Yukawa and QCD radiative corrections, are fixed by 
$m_{H^+},\tan\b,\mu, m_t,m_b,M_{\ti Q},M_{\ti U},M_{\ti D},A_t,A_b,
|M_1|,M_2$, and $m_{\sg}$ \cite{ref3}. The neutral Higgs mass eigenstates 
$H_1^0, H_2^0$ and $H_3^0$ are mixtures of CP-even and CP-odd states 
($\phi_{1,2}$ and $a$) due to the explicit CP violation in the Higgs sector. 
For the radiatively corrected masses and mixings of the Higgs bosons we use 
the formulae of Ref.\cite{ref3}. 

Here we list possible important decay modes of $\st_{1,2}$ and $\sb_{1,2}$: 
\beqaa
  \st_1 & \to & t \sg \, , t \nt_i \, , \, b \chp_j \, , \, 
                \sb_1 W^+ \, , \, \sb_1 H^+
                                                           \label{eq:i} \\
  \st_2 & \to & t \sg \, , t \nt_i \, , \, b \chp_j \, , \,
                \st_1 Z^0 \, , \, \sb_{1,2} W^+ \, , \, \st_1 H_k^0 \, , \, 
                \sb_{1,2} H^+ 
                                                           \label{eq:j} \\
  \sb_1 & \to & b \sg \, , b \nt_i \, , \, t \chm_j \, , \, 
                \st_1 W^- \, , \, \st_1 H^-
                                                           \label{eq:k} \\
  \sb_2 & \to & b \sg \, , b \nt_i \, , \, t \chm_j \, , \,
                \sb_1 Z^0 \, , \, \st_{1,2} W^- \, , \, \sb_1 H_k^0 \, , \, 
                \st_{1,2} H^-. 
                                                           \label{eq:l} 
\eeqaa
The decays into a gauge or Higgs boson in (\ref{eq:i})-(\ref{eq:l}) 
are possible in case the mass splitting between the squarks is sufficiently 
large \cite{stop2-stau2, stop1}. The explicit expressions of the widths of the 
decays (\ref{eq:i})-(\ref{eq:l}) in case of real SUSY parameters are given in 
\cite{BMP}. Those for complex parameters can be obtained by using the 
corresponding masses and couplings (mixings) from 
Refs.\cite{ref3,ref7,BMP} and will be presented elsewhere 
\cite{long_CPsq_paper}.

The phase dependence of the widths stems from that of the involved 
mass-eigenvalues, mixings and couplings among the interaction-eigenfields. 
Here we summarize the most important features of the phase dependences.
\renewcommand{\labelenumi}{(\Roman{enumi})} 
\begin{enumerate}
  \item $\sq_i$  ($\sq = \st, \sb$) sectors:
    \begin{enumerate}
      \item The mass-eigenvalues $m_{\sq_{1,2}}$ are sensitive to the phases 
            $(\varphi_{A_q}, \varphi_{\mu})$ 
            via $\cos(\varphi_{A_q} + \varphi_{\mu})$
            if and only if $|a_q m_q| \sim (m_{\sq_L}^2 + m_{\sq_R}^2)/2$ 
            {\em and} $|A_q| \sim |\mu| C_q$ (with $C_t = \cot\b$ and 
            $C_b = \tan\b$).
      \item The $\sq$-mixing angle $\t_{\sq}$ (given by 
            $\tan 2\t_{\sq} = 2|a_q m_q|/(m_{\sq_L}^2 - m_{\sq_R}^2)$ ) 
            is sensitive to $(\varphi_{A_q}, \varphi_{\mu})$ via 
            $\cos(\varphi_{A_q} + \varphi_{\mu})$  if and only if 
            $2|a_q m_q| \gsim |m_{\sq_L}^2 - m_{\sq_R}^2|$ {\em and} 
            $|A_q| \sim |\mu| C_q$. 
      \item The $\sq$-mixing phase $\varphi_{\sq}$ in Eq.(\ref{eq:e}) is 
            sensitive to 
            ($\varphi_{A_q}$, $\varphi_{\mu}$), $\varphi_{A_q}$ and 
            $\varphi_{\mu}$ if $|A_q| \sim |\mu| C_q$, $|A_q| \gg |\mu| C_q$ 
            and $|A_q| \ll |\mu| C_q$, respectively. 
            For large squark mixing the term
            $\propto \sin 2\theta_{\tilde{q}} 
            \cos\varphi_{\tilde{q}}$ can result in a
            large phase dependence of the decay widths \cite{long_CPsq_paper} 
            (see Eq.(\ref{eq:f})).
    \end{enumerate}
  \item $\nt_i$ and $\ch_j$ sectors:
    \begin{enumerate}
      \item $m_{\nt_i}$ ($i=1,...,4$) and the $\nt$-mixing matrix are sensitive 
            $[$insensitive$]$ to the phases $(\varphi_1, \varphi_{\mu})$ 
            for small $[$large$]$ $\tan\b$.
      \item $m_{\ch_{1,2}}$ and the $\ch$-mixing matrices are sensitive 
            $[$insensitive$]$ to $\varphi_{\mu}$ for small $[$large$]$ 
            $\tan\b$.
    \end{enumerate}
  \item $H^\pm$ sector:\\
        This sector is independent of the phases. 
  \item $H_i^0$ sector:
    \begin{enumerate}
      \item $m_{H_i^0}$ ($i=1,2,3$) are sensitive $[$insensitive$]$ to the 
            phase sums $\varphi_{A_{t,b}} + \varphi_{\mu}$ for small 
            $[$large$]$ $\tan\b$ \cite{ref3}.
      \item In general the $H_i^0$-mixing matrix (a real orthogonal 
            $3 \times 3$ matrix $O_{ij}$) is sensitive to 
            $\varphi_{A_{t,b}} + \varphi_{\mu}$ for any $\tan\b$ \cite{ref3}.
    \end{enumerate}
  \item The couplings among the interaction-eigenfields:
    \begin{enumerate}
      \item For the decays into fermions and gauge bosons in 
            Eqs.(\ref{eq:i})-(\ref{eq:l}), they are gauge couplings and$/$or 
            Yukawa couplings ($h_{t,b}$), and are independent of 
            the phases at tree-level. 
      \item For the decays into Higgs bosons in Eqs.(\ref{eq:i})-(\ref{eq:l}), 
            the $\sq_L$-$\sq_R$ ($\sq_L$-$\sq_L$, $\sq_R$-$\sq_R$) couplings are 
            dependent on (independent of) the phases $\varphi_{A_t}$, 
            $\varphi_{A_b}$ and $\varphi_{\mu}$;
            \beqaa
              C(\sb_L^\dagger \st_R H^-) & \sim & \sin\b \, h_t 
                                              (A_t^* \cot\b + \mu) 
                                                           \label{eq:m} \\
              C(\st_L^\dagger \sb_R H^+) & \sim & \cos\b \, h_b 
                                              (A_b^* \tan\b + \mu) 
                                                           \label{eq:n} \\
              C(\st_L^\dagger \st_R \phi_1) & \sim & h_t \mu 
                                                           \label{eq:o} \\
              C(\st_L^\dagger \st_R \phi_2) & \sim & h_t A_t^* 
                                                           \label{eq:p} \\
              C(\st_L^\dagger \st_R a) & \sim & \sin\b \, h_t 
                                              (A_t^* \cot\b + \mu) 
                                                           \label{eq:q} \\
              C(\sb_L^\dagger \sb_R \phi_1) & \sim & h_b A_b^* 
                                                           \label{eq:r} \\
              C(\sb_L^\dagger \sb_R \phi_2) & \sim & h_b \mu 
                                                           \label{eq:s} \\
              C(\sb_L^\dagger \sb_R a) & \sim & \cos\b \, h_b 
                                              (A_b^* \tan\b + \mu) 
                                                           \label{eq:t} 
            \eeqaa
            with 
            \beqaa
              h_t &=& g \, m_t / (\sqrt{2} m_W \sin\b) 
                                                           \label{eq:u} \\
              h_b &=& g \, m_b / (\sqrt{2} m_W \cos\b). 
                                                           \label{eq:v} 
            \eeqaa
            Here $\phi_i = O_{ij} H_j^0$ ($i=1,2$) and $a = O_{3j} H_j^0$ 
            are the CP-even and CP-odd neutral Higgs bosons, respectively 
            \cite{ref3}. 
    \end{enumerate}
\end{enumerate}
According to items (I)-(V) we expect that the widths 
(and hence the branching ratios) of the 
decays (\ref{eq:i})-(\ref{eq:l}) are sensitive to the phases 
($\varphi_{A_t}$, $\varphi_{A_b}$, $\varphi_{\mu}$, $\varphi_1$) in a large 
region of the MSSM parameter space. 

\section{Numerical Results}

Now we turn to the numerical analysis of the $\st_{1,2}$ and $\sb_{1,2}$ 
decay branching ratios. We calculate the widths of all possibly 
important two-body decay modes of Eqs.(\ref{eq:i})-(\ref{eq:l}). 
Three-body decays are negligible in the parameter space under study. 
In order to improve the convergence of the perturbative expansion 
\cite{stop1,improvedQCDcorr} we calculate the 
tree-level widths by using the corresponding tree-level couplings defined 
in terms of ``effective" MSSM running quark masses $m_{t,b}^{run}$ 
(i.e. those defined in terms of 
the effective running Yukawa couplings $h_{t,b}^{run}$ 
$\propto m_{t,b}^{run}$). For the kinematics, e.g., for 
the phase space factor we use the on-shell masses obtained by using the 
on-shell (pole) quark masses $M_{t,b}$. 
We take $M_t=175$ GeV, $M_b=5$ GeV, $m_t^{run}=150$ GeV, $m_b^{run}=3$ GeV, 
$m_Z=91.2$ GeV, $\sin^2\t_W =0.23$, $m_W = m_Z \cos\t_W$, 
$\a(m_Z)=1/129$, and $\alpha_s(m_Z)=0.12$ (with 
$\alpha_s(Q)=12\pi/((33-2n_f)\ln(Q^2/\Lambda_{n_f}^2))$, $n_f$ being the 
number of quark flavors). In order not to vary too many parameters we fix 
$|A_t|=|A_b| \equiv |A|$ and $M_2=300$ GeV, i.e. $m_{\sg}=820$ GeV. 
In the following we will assume that $m_{\sg}>m_{\st_2,\sb_2}$ so that  
the decays $\tilde t_i \to t \tilde g$ and $\tilde b_i \to b \tilde g$ 
are kinematically forbidden.
In our numerical study 
we take $\tan\b$, $m_{\st_1}$, $m_{\st_2}$, $m_{\sb_1}$, $|A|$, $|\mu|$, 
$\varphi_{A_t}$, $\varphi_{A_b}$, $\varphi_{\mu}$, $\varphi_1$ and 
$m_{H^+}$ as input parameters, where $m_{\st_{1,2}}$ and $m_{\sb_1}$ are 
the on-shell squark masses.  
Note that for a given set of the input parameters we 
obtain two solutions for ($M_{\ti Q}$, $M_{\ti U}$) corresponding to 
the two cases $m_{\st_L} \geq m_{\st_R}$ and $m_{\st_L} < m_{\st_R}$ 
from Eqs. (\ref{eq:a})-(\ref{eq:d}) and (\ref{eq:g}) with $m_t$ replaced 
by $M_t$. In the plots we impose the following conditions in order to 
respect experimental and theoretical constraints:
\renewcommand{\labelenumi}{(\roman{enumi})} 
\begin{enumerate}
  \item $m_{\ch_1} > 103$ GeV, $m_{\nt_1} > 50$ GeV,
        $m_{\st_1,\sb_1} > 100$ GeV, 
        $m_{\st_1,\sb_1} > m_{\nt_1}$, \\
        $m_{H_1^0} > 105$ GeV,  
  \item $|A_t|^2 < 3\,(M_{\ti Q}^2 + M_{\ti U}^2 + m_2^2)$, and 
        $|A_b|^2 < 3\,(M_{\ti Q}^2 + M_{\ti D}^2 + m_1^2)$, where 
        $m_1^2=(m_{H^+}^2 + m_Z^2 \sin^2\t_W)\sin^2\b-\frac{1}{2}\,m_Z^2$ and 
        $m_2^2=(m_{H^+}^2 + m_Z^2 \sin^2\t_W)\cos^2\b-\frac{1}{2}\,m_Z^2$, 
  \item $\Delta\rho\,(\st \!-\! \sb) < 0.0012$ 
        \cite{ref10} using the formula of \cite{ref11}, 
  \item $2.0 \cdot 10^{-4} < B(b \to s \gamma) < 4.5 \cdot 10^{-4}$ 
        \cite{bsgamma} assuming the Kobayashi-Maskawa mixing also 
        for the squark sector.
\end{enumerate}
Condition (i) is imposed to satisfy the experimental mass bounds from LEP 
\cite{LEP}. (ii) is the approximate necessary condition for the tree-level 
vacuum stability \cite{Derendinger-Savoi}. (iii) constrains $\mu$ and $\tan\b$ 
(in the squark sector). 
For the calculation of the $b\to s\gamma$ width we use the formula
of \cite{Bertolini:1990if} including the O($\alpha_s$) corrections as given 
in \cite{Kagan:1998ym}.


\noi
As alreaday mentioned the experimental upper limits 
on the EDMs of electron, neutron, $^{199}$Hg and $^{205}$Tl
strongly constrain the SUSY CP phases \cite{Nath}. 
One interesting possibility for evading these constraints is to invoke large 
masses (much above the TeV scale) for the first two generations of the 
sfermions \cite{Nelson}, keeping the third generation sfermions relatively 
light ($\lsim$1 TeV). In such a scenario ($\varphi_1$, $\varphi_{\mu}$) and 
the CP phases of the third generation ($\varphi_{A_t}$, $\varphi_{A_b}$, 
$\varphi_{A_\tau}$) are practically unconstrained \cite{Nelson}. We adopt this 
scenario. 
Furthermore, we have checked that the electron and neutron 
EDM constraints at two-loop 
level \cite{Pilaftsis} are fulfilled in the numerical examples studied 
in this article.

In Fig.1 we plot the contours of the branching ratios of the $\st_1$ decays 
$B(\st_1 \to t \nt_1)$ and $B(\st_1 \to b \chp_1)$ in the 
$|A|-|\mu|$ plane for $\varphi_{A_t}=0$ and $\pi/2$ with $\tan\b=8$, 
($m_{\st_1}$,$m_{\st_2}$,$m_{\sb_1}$) = (400,700,200) GeV, 
$\varphi_{A_b}=\varphi_1=0$, $\varphi_{\mu}=\pi$, and $m_{H^+}=600$ GeV
in the case $m_{\st_L} \geq m_{\st_R}$. In the case 
$m_{\st_L} < m_{\st_R}$ we have obtained similar results. 
As expected, these branching ratios are sensitive to the phase 
$\varphi_{A_t}$ in a sizable region of the $|A|-|\mu|$ plane. 
The difference between $\varphi_{A_t}=0$
(Figs. 1a and c) and $\varphi_{A_t}=\pi/2$
(Figs. 1b and d) can be explained by item (I).
Especially the strong $\varphi_{A_t}$ dependence of the decay
width $\Gamma(\tilde t_1 \to b \tilde\chi^+_1)$ is
a result of item (I)(c) \cite{Bartl:2003yp}.
Moreover, the $\st$-mixing angle $\tst$ is sensitive to 
$\cos(\varphi_{A_t}+\varphi_{\mu})$ for large $|A_t|$(= $|A|$) with 
$|A_t| \sim |\mu|/8$, which is a consequence of item (I)(b).

In Fig.2 we plot the contours of the $\st_1$ decay branching ratios 
$B(\st_1 \to t \nt_1)$ and $B(\st_1 \to b \chp_1)$ in the 
$\varphi_{A_t}-\varphi_{\mu}$ plane for $\tan\b=8$, ($m_{\st_1}$,$m_{\st_2}$,
$m_{\sb_1}$)=(400,700,200) GeV, $|A|=800$ GeV, $|\mu|=500$ GeV, 
 $\varphi_{A_b}$=$\varphi_1$=0, and $m_{H^+}=600$ GeV 
in the case $m_{\st_L} \geq m_{\st_R}$. In the case 
$m_{\st_L} < m_{\st_R}$ we have obtained similar results. 
One sees that these branching ratios depend quite strongly on the CP phases 
$\varphi_{A_t}$ and $\varphi_{\mu}$, as follows 
from item (I). 

In Fig.3 we show the $\varphi_{\mu}$ dependence of the $\st_1$ decay 
branching ratios $B(\st_1 \to t \nt_1)$ and $B(\st_1 \to b \chp_1)$ for 
$\varphi_1=0$ and $\pi/2$ with $\tan\b=5$, ($m_{\st_1}$,$m_{\st_2}$,
$m_{\sb_1}$)=(400,700,200) GeV, $|A|=800$ GeV, $|\mu|=500$ GeV, 
$\varphi_{A_t}=\varphi_{A_b}=0$, and $m_{H^+}=600$ GeV in the case 
$m_{\st_L} \geq m_{\st_R}$. For $m_{\st_L} < m_{\st_R}$ 
we have obtained similar results.
As can be seen in Fig.3 the branching ratios are very [somewhat] sensitive to 
$\varphi_{\mu}$ $[\varphi_1]$ for small $\tan\b$=5. 
%
%
For this choice of parameters the masses and 
mixings of $\st$, $\sb$, $\nt$ and $\ch$ are sensitive to $\varphi_{\mu}$, 
while only the masses and mixings of $\nt$ are sensitive to $\varphi_1$ ,
as can be seen from items (I) and (II). 
In general, according to items (I), (II) and (V), the $\st_1$ decay 
branching ratios tend to be sensitive [insensitive] to $\varphi_{A_t}$ 
[$\varphi_{A_b}$, $\varphi_{\mu}$, and $\varphi_1$] 
for large $\tan\b (\gsim 15)$ in case the width of the $\sb_1 H^+$ mode in Eq. 
(\ref{eq:i}) (which can be sensitive to ($\varphi_{A_t}$, $\varphi_{A_b}$) 
as seen from Eqs. (\ref{eq:f}, \ref{eq:m}, \ref{eq:n})) is relatively small.
We have confirmed this.

In Fig.4 we show the $\varphi_{A_t}$ dependence of the $\st_2$ decay 
branching ratios for $\varphi_\mu=0$ and $\pi/2$ with $\tan\b$=8, 
($m_{\st_1}$,$m_{\st_2}$,$m_{\sb_1}$)=(400,700,200) GeV, $|A|=800$ GeV, 
$|\mu|=500$ GeV, $\varphi_{A_b}$=$\varphi_1$=0, and $m_{H^+}=600$ GeV 
in the case $m_{\st_L} \geq m_{\st_R}$. The case $m_{\st_L} < m_{\st_R}$ 
leads to similar results. We see that the $\st_2$ decay branching 
ratios are very sensitive to $\varphi_{A_t}$ and depend significantly on 
$\varphi_{\mu}$. This can be expected from items (I), (II), (IV) and (V)(b). 
%
%
From item (I)(c) [(IV)(b) and (V)(b)] we expect that the widths of the decays 
$\st_2 \to b \chp_{1,2}$ [$\st_1 H_k^0$ and $\sb_{1,2} H^+$] in Eq.(\ref{eq:j}) 
can be sensitive to $\varphi_{A_t}$ [$\varphi_{A_{t,b}}$ and $\varphi_{\mu}$] 
for large $\tan\b (\gsim 15)$ if they are kinematically allowed. 
We have confirmed this.
%

For $\sb_{1,2}$ decays we have obtained results similar to those for the 
$\st_{1,2}$ decays. Here we show just a few typical 
results for them. In Fig.5 we show the $\varphi_{A_b}$ dependence of the 
$\sb_1$ [$\sb_2$] decay branching ratios for $\tan\b$=30, ($m_{\st_1}$,
$m_{\st_2}$, $m_{\sb_1}$)=(200,700,400) GeV [(175,500,350) GeV], 
($|A|$, $|\mu|$)=(800,700) GeV [(600,500) GeV], 
$\varphi_{A_t}$=$\varphi_{\mu}$=$\pi$, $\varphi_1$=0, and $m_{H^+}=180$ GeV 
in the case $m_{\st_L} \geq m_{\st_R}$. For $m_{\st_L} < m_{\st_R}$ 
our results are similar. 
We find that the $\sb_{1,2}$ decay branching ratios 
are very sensitive to $\varphi_{A_b}$ as expected from items (IV)(b) 
and (V)(b). 
The main reason is that the decay widths for
$\tilde b_{1,2}\to \tilde t_1 H^-$ strongly  
depend on $\varphi_{A_b}$ (and $\varphi_{\mu}$)
for large $\tan\beta$ (see Eqs.(\ref{eq:m},\ref{eq:n})). This explains
also the tendency of the $\varphi_{A_b}$ dependence of the branching 
ratios for the other decays shown in Fig.5. 
For small $\tan\beta\sim 8$ we expect that
the $\tilde b_{1,2}$ decay branching ratios
can be somewhat sensitive to $\varphi_{A_{t,b}}$
and $\varphi_{\mu}$ (see items (I) and (V)(b)).
Similarly, we expect that the decay widths of 
$\tilde t_{1,2}H^-$ can be fairly sensitive
to $\varphi_{A_{t,b}}$ and $\varphi_{\mu}$.
We have confirmed this.


\section{Conclusions}

We have calculated the branching ratios of the two-body
decays of $\tilde{t}_{1,2}$ and $\tilde{b}_{1,2}$ and studied their CP 
phase dependence within the MSSM with complex parameters $A_t$, $A_b$, 
$\mu$ and $M_1$. We have shown that the effect of the SUSY CP phases
on the branching ratios can be quite strong in a large domain of the
MSSM parameter space. 
The $\varphi_{A_b}$ dependence of 
the $\tilde b_{1,2}$ decay branching
ratios is mainly due to the $\varphi_{A_b}$ dependence of 
their couplings to the Higgs bosons.
In the case of the $\tilde{t}_{1,2}$ decays the 
branching ratios depend on $\varphi_{A_t}$ also via
the $\tilde t$-mixing phase 
$\varphi_{\tilde t} \approx \varphi_{A_t}$ for 
$|A_t|\gg |\mu|/\tan\beta$, in addition to
the $\varphi_{A_t}$ dependence of the
mixing angle $\theta_{\tilde t}$ and their couplings
to the Higgs bosons. Some of the branching ratios 
can change by a factor of 2 or more when varying the phases.
This CP phase dependence of the branching ratios 
could have an important impact on the search
for $\tilde{t}_{1,2}$ and $\tilde{b}_{1,2}$ and on the determination
of the MSSM parameters at future colliders.
%

\section*{Acknowledgements}
The authors thank W. Majerotto and Y. Yamada for helpful discussions.
W.P. thanks E. Lunghi for useful discussions on the $b\to s\gamma$
constraint.
This work was supported by the `Fonds zur F\"orderung der wissenschaftlichen 
Forschung' of Austria, Projects No. P13139-PHY and No. P16592-N02
, and by the European Community's 
Human Potential Programme under contract HPRN-CT-2000-00149. 
W.P.~is supported by `Fonds zur F\"orderung der wissenschaftlichen 
Forschung' of Austria, Erwin Schr\"odinger fellowship Nr. J2272, and 
partly by the `Schweizer Nationalfonds'. 


\newpage



\begin{flushleft}
{\Large \bf Figure Captions} \\
\end{flushleft}

\noi
{\bf Figure 1}: 
Contours of the $\st_1$ decay branching ratios $B(\st_1 \to t \nt_1)$ (a,b) 
and $B(\st_1 \to b \chp_1)$ (c,d) in the $|A|-|\mu|$ plane 
for $\varphi_{A_t}=0$ (a,c) and $\pi/2$ (b,d) with $\tan\b=8$, 
($m_{\st_1}$,$m_{\st_2}$,$m_{\sb_1}$) = (400,700,200) GeV, 
$\varphi_{A_b}$=$\varphi_1$=0, $\varphi_{\mu}$=$\pi$, and $m_{H^+}=600$ GeV 
in the case $m_{\st_L} \geq m_{\st_R}$. 
The blank areas are excluded by the conditions (i) to (iv) 
given in the text.

\noi 
{\bf Figure 2}: 
Contours of the $\st_1$ decay branching ratios $B(\st_1 \to t \nt_1)$ (a) 
and $B(\st_1 \to b \chp_1)$ (b) in the $\varphi_{A_t}$-$\varphi_{\mu}$ 
plane for $\tan\b=8$, ($m_{\st_1}$,$m_{\st_2}$,$m_{\sb_1}$)=
(400,700,200) GeV, $|A|=800$ GeV, $|\mu|=500$ GeV, 
 $\varphi_{A_b}$=$\varphi_1$=0, and $m_{H^+}=600$ GeV 
in the case $m_{\st_L} \geq m_{\st_R}$.

\noi 
{\bf Figure 3}: 
$\varphi_{\mu}$ dependence of the $\st_1$ decay branching ratios 
$B(\st_1 \to t \nt_1)$ and $B(\st_1 \to b \chp_1)$ 
for $\varphi_1=0$ (solid lines) and $\pi/2$ (dashed lines) with $\tan\b=5$, 
($m_{\st_1}$,$m_{\st_2}$,$m_{\sb_1}$)=(400,700,200) GeV, $|A|=800$ GeV, 
$|\mu|=500$ GeV, $\varphi_{A_t}$=$\varphi_{A_b}$=0, and $m_{H^+}=600$ GeV 
in the case $m_{\st_L} \geq m_{\st_R}$.

\noi 
{\bf Figure 4}: 
$\varphi_{A_t}$ dependence of the $\st_2$ decay 
branching ratios for $\varphi_\mu=0$ (a) and $\pi/2$ (b) with $\tan\b$=8, 
($m_{\st_1}$,$m_{\st_2}$,$m_{\sb_1}$)=(400,700,200) GeV, $|A|=800$ GeV, 
$|\mu|=500$ GeV, $\varphi_{A_b}$=$\varphi_1$=0, and $m_{H^+}=600$ GeV 
in the case $m_{\st_L} \geq m_{\st_R}$. Note that the $\st_1 H_{2,3}^0$ 
and $\sb_1 H^+$ modes are kinematically forbidden here.

\noi 
{\bf Figure 5}: 
$\varphi_{A_b}$ dependence of the $\sb_1$ (a) [$\sb_2$ (b)] decay 
branching ratios for $\tan\b$=30, ($m_{\st_1}$,$m_{\st_2}$,$m_{\sb_1}$)=
(200,700,400) GeV [(175,500,350) GeV], ($|A|$, $|\mu|$)=(800,700) GeV 
[(600,500) GeV], $\varphi_{A_t}$=$\varphi_{\mu}$=$\pi$, $\varphi_1$=0, 
and $m_{H^+}=180$ GeV in the case $m_{\st_L} \geq m_{\st_R}$. 
Only interesting modes are shown in Fig.b where we have 
$m_{\sb_2} \sim $ 570GeV and 
($m_{H_1^0}$,$m_{H_2^0}$,$m_{H_3^0}$) $\sim$ (114,156,158) GeV. 

\newpage
%
%
%
\begin{figure}[!htb] 
\begin{center}
\scalebox{0.55}[0.66]{\includegraphics{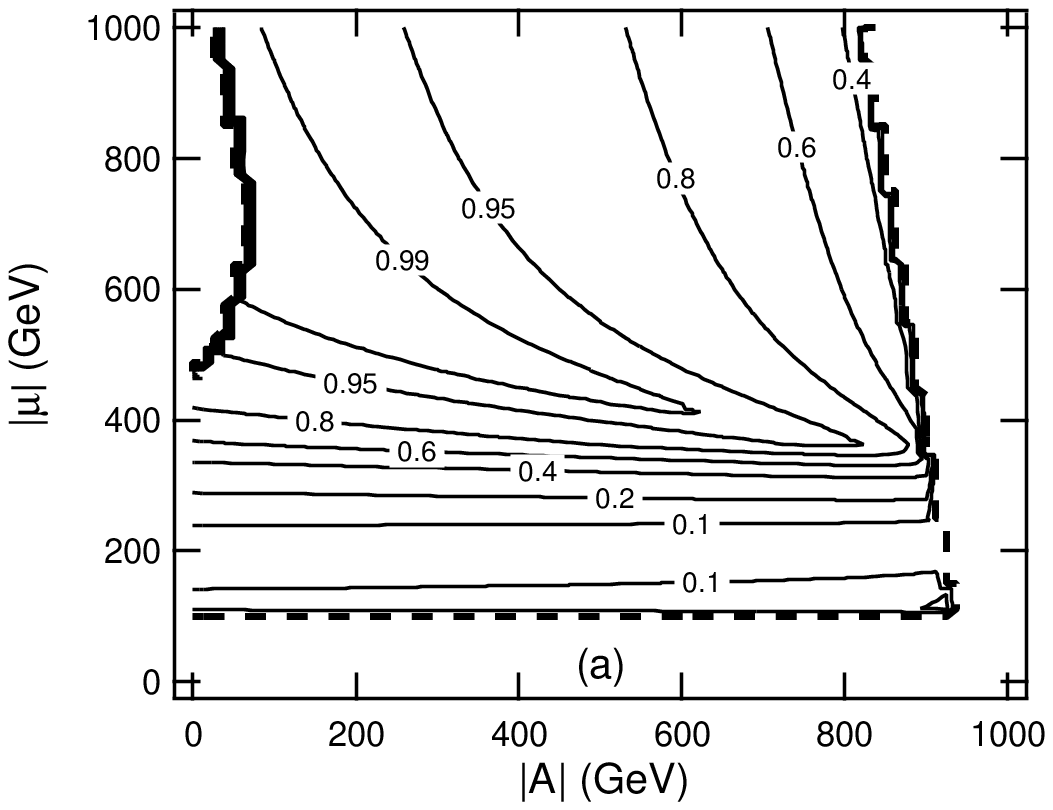}} \hspace{-10.0mm} 
\scalebox{0.55}[0.66]{\includegraphics{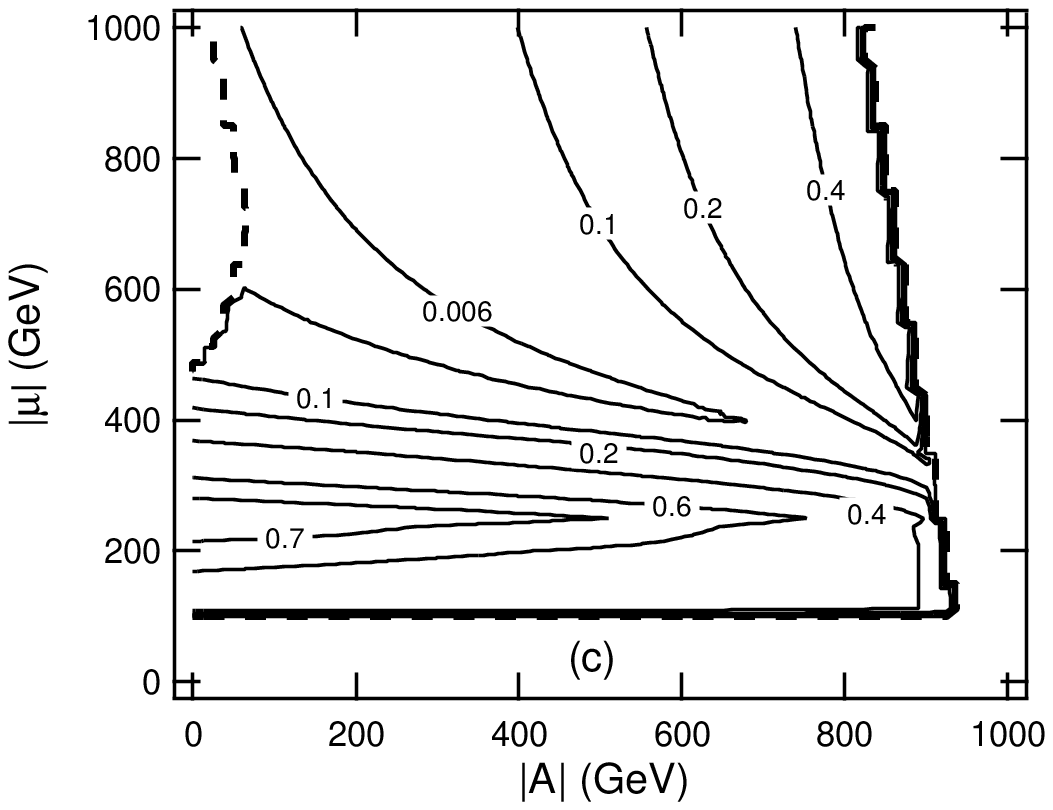}} \\ 
\scalebox{0.55}[0.66]{\includegraphics{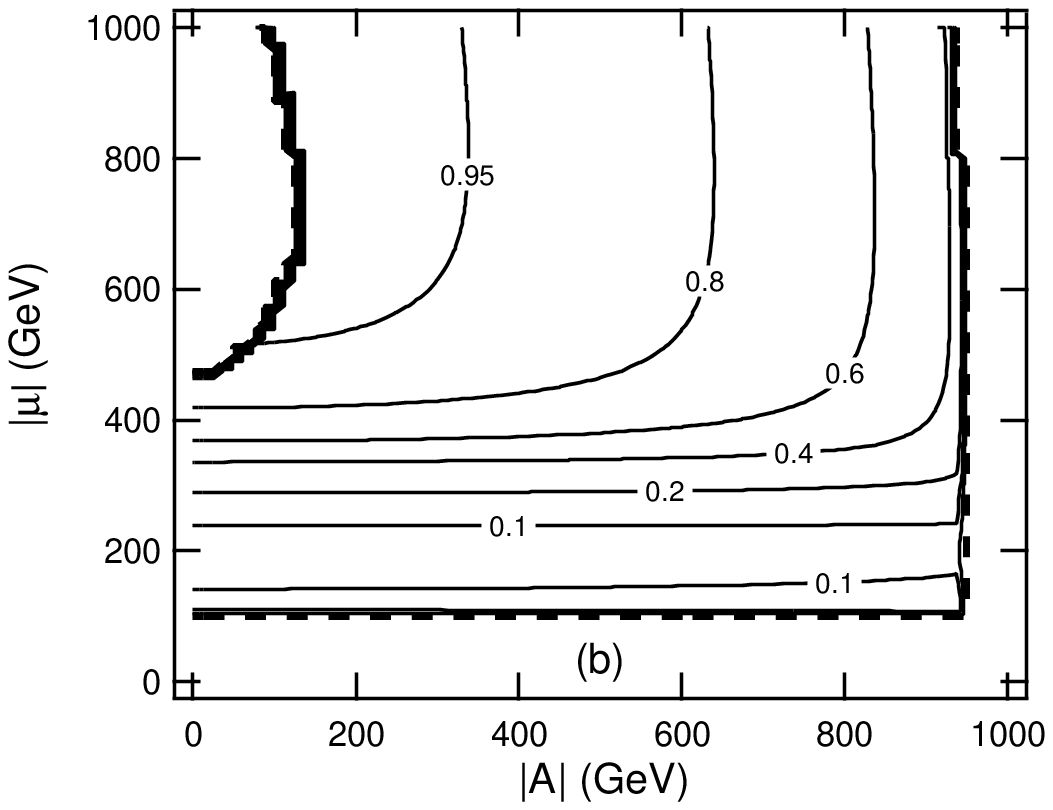}} \hspace{-10.0mm} 
\scalebox{0.55}[0.66]{\includegraphics{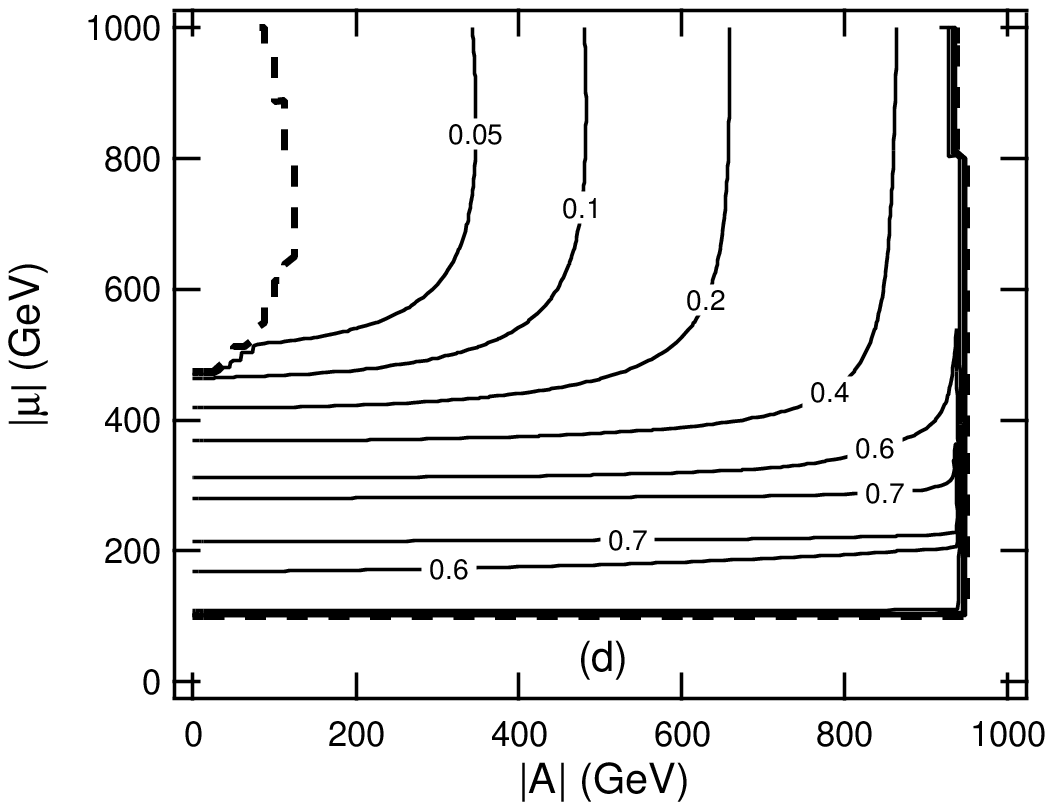}} \\ 
\vspace{10mm}
{\LARGE \bf Fig.1}
\end{center}
\end{figure}
%

\newpage
%
%
\begin{figure}[!htb] 
\begin{center}
\scalebox{0.6}[0.8]{\includegraphics{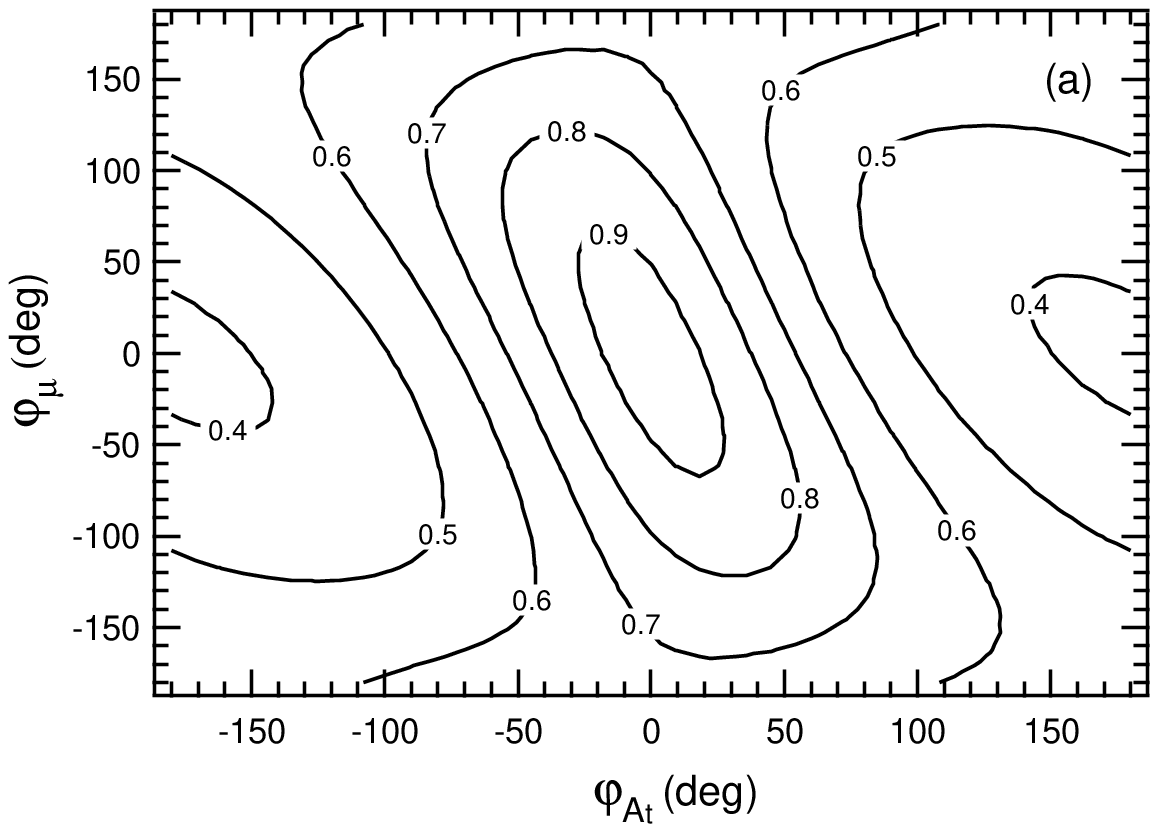}} \\ 
\scalebox{0.6}[0.8]{\includegraphics{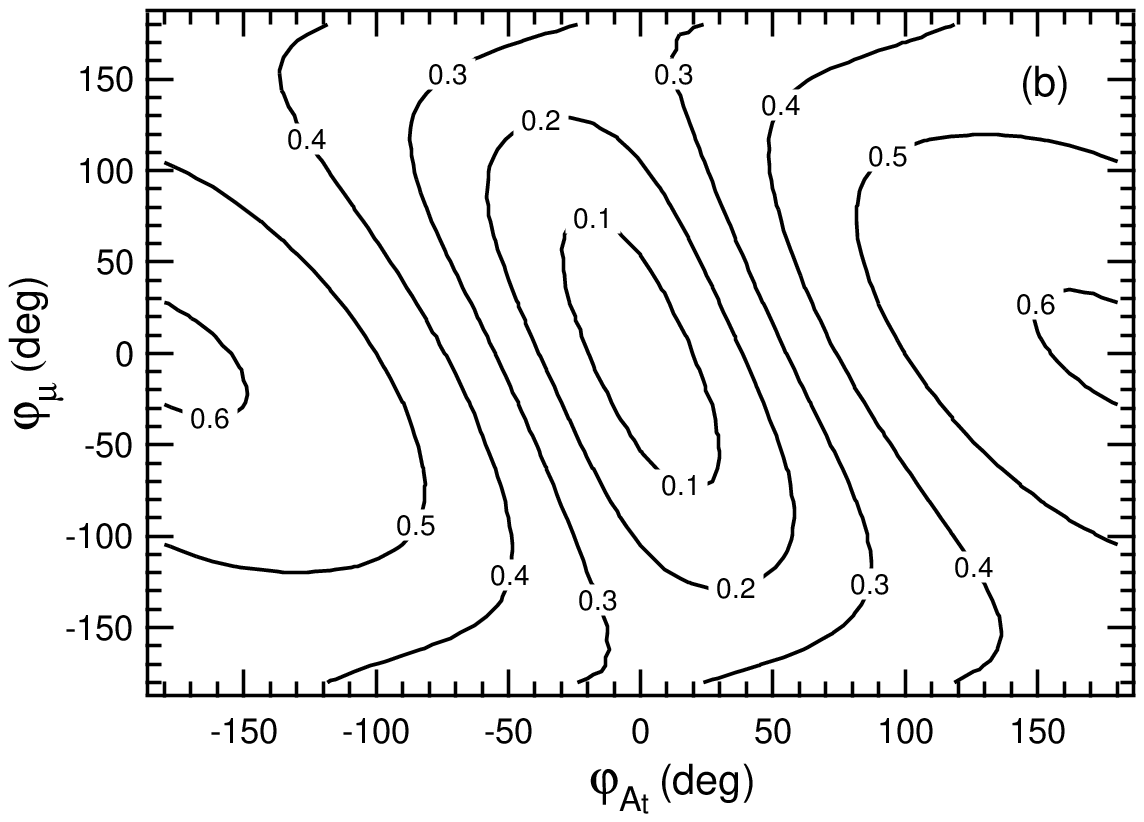}} \\ 
\end{center}
\end{figure}

\begin{center}
{\LARGE \bf Fig.2}
\end{center}

\newpage
%
%
\begin{figure}[!htb] 
\begin{center}
\scalebox{0.8}[1.2]{\includegraphics{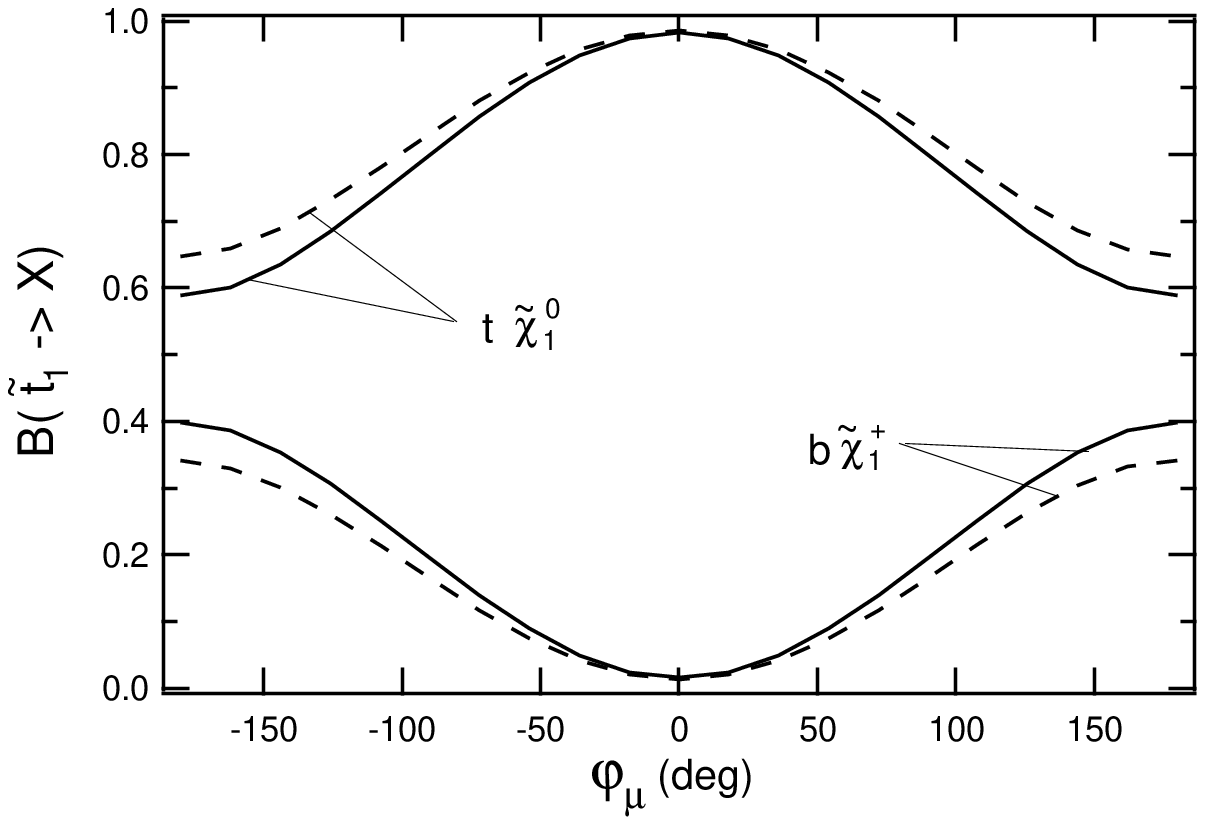}} 
\end{center}
\end{figure}

\begin{center}
{\LARGE \bf Fig.3}
\end{center}

\newpage
%
%
\begin{figure}[!htb] 
\begin{center}
\scalebox{0.7}[1.1]{\includegraphics{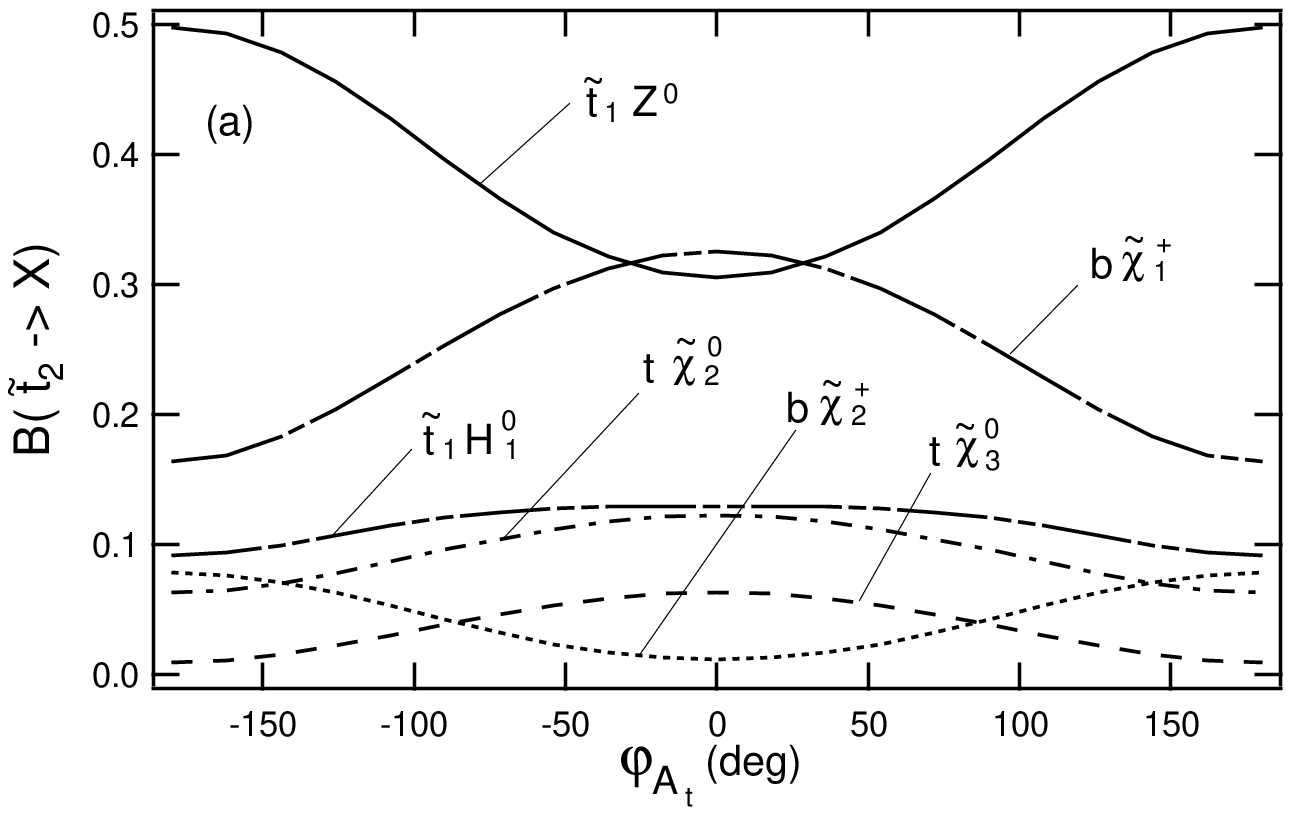}} \\ 
\vspace{-10mm}
\scalebox{0.7}[1.1]{\includegraphics{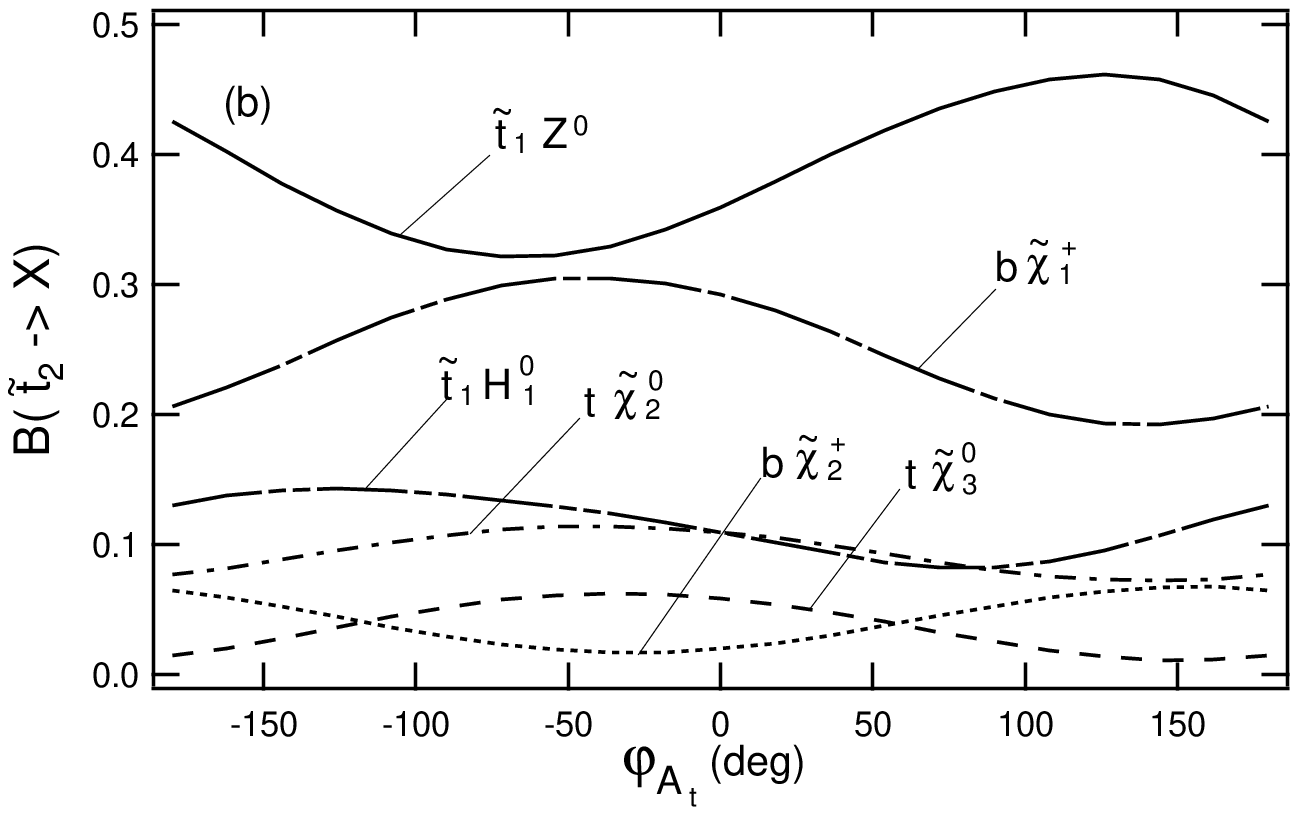}} \\ 
\vspace{5mm}
{\LARGE \bf Fig.4}
\end{center}
\end{figure}
%

\newpage
%
%
\begin{figure}[!htb] 
\begin{center}
\scalebox{0.8}[1.1]{\includegraphics{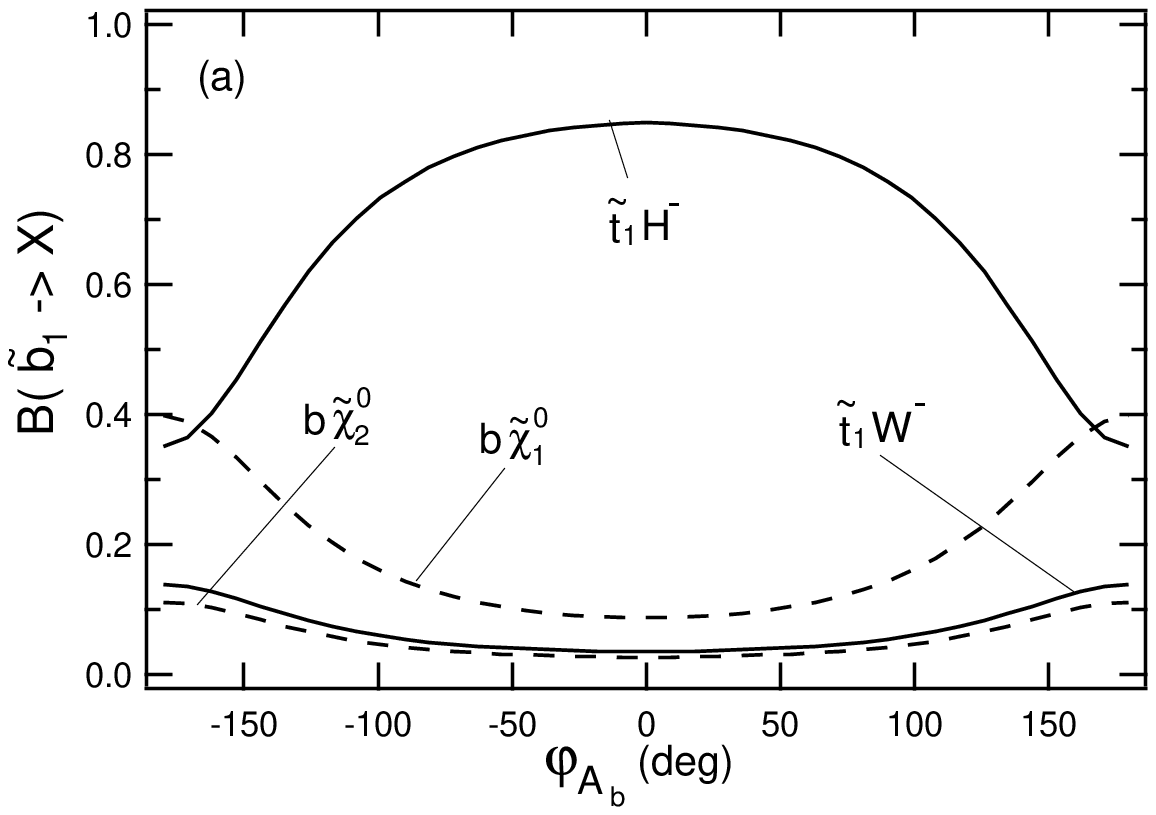}} \\ 
\vspace{-7mm}
\scalebox{0.8}[1.1]{\includegraphics{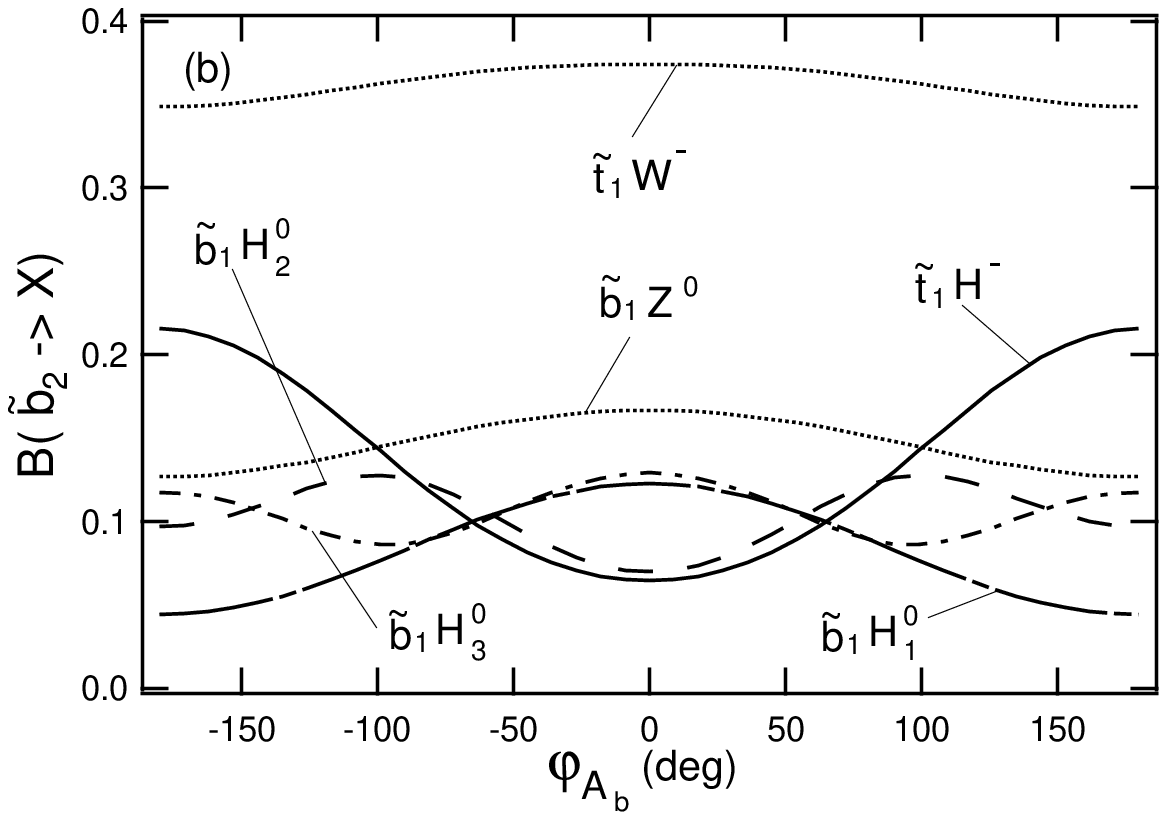}} \\ 
\vspace{5mm}
{\LARGE \bf Fig.5}
\end{center}
\end{figure}
%


\end{document}